\documentclass[11pt,a4paper]{article}

\usepackage[T1]{fontenc}
\usepackage[margin=1in]{geometry}
\usepackage{amsmath,amssymb,amsthm}
\usepackage[protrusion=true,expansion=false]{microtype}
\usepackage{booktabs}
\usepackage{tabularx}
\usepackage[hidelinks]{hyperref}
\usepackage{cleveref}
\usepackage{enumitem}
\usepackage{xcolor}
\usepackage{natbib}

\newcommand{\Topen}{T_{\text{open}}}
\newcommand{\Tnews}{T_{\text{news}}}
\newcommand{\Tevent}{T_{\text{event}}}
\newcommand{\Tres}{T_{\text{resolve}}}
\newcommand{\Tdead}{T_{\text{deadline}}}
\newcommand{\ILS}{\mathrm{ILS}}

\title{\textbf{Per-Market Information Leakage and Order-Flow Skill:\\ Two Methodological Lenses on Informed Trading\\ in Decentralized Prediction Markets}}

\author{Maksym Nechepurenko\thanks{Research Department, Devnull FZCO, Dubai, UAE. Email: \texttt{maksym@devnull.ae}.}}

\date{Preprint v2 (revised) --- \today}

\begin{document}

\maketitle

\begin{abstract}
\noindent
April 2026 saw notable methodological convergence in the academic study of informed trading on decentralized prediction markets. Three methodological approaches surfaced almost simultaneously, each addressing the same body of documented cases on Polymarket but operating at different methodological layers: \citet{mitts2026iran} apply a composite statistical screen to over $210{,}000$ wallet--market pairs and estimate \$$143$ million in aggregate anomalous profit; \citet{gomezcram2026informed} apply an event-level sign-randomization test to the platform's complete transaction history, classifying $3.14\%$ of accounts as ``skilled winners'' who drive most price discovery, and separately apply a single-event lifecycle-and-conviction heuristic that flags $1{,}950$ accounts as ``insiders''; \citet{nechepurenko2026foresightflow_methodology, nechepurenko2026foresightflow_empirical} develop the \emph{Information Leakage Score} (ILS) framework, which quantifies per-market information front-loading at the article-derived public-event timestamp.

This paper provides a methodological comparison and a sketch of how the approaches combine. The central organizing claim is that these are three distinct \emph{layers} of detection, not three competing methods on a single layer. Sign-randomization, in particular, is best understood as an account-level test of \emph{persistent directional skill conditional on opportunity selection}---not a direct test of insider trading, and not a per-market measure. The heuristic insider flag in \citet{gomezcram2026informed} is methodologically separate from their skill classifier, applies to a population (single-event, recently-created accounts) that the skill classifier explicitly excludes by design, and has unknown precision against an external labelled set. The Polymarket sample on which all three approaches are evaluated pools politics, sports, crypto, and other categories with structurally different information technologies, so a platform-wide ``skilled winner'' classification is mechanism-ambiguous and category-conditioned decompositions are required before the methodology can be used as a surveillance layer.

Against this backdrop, ILS$^{\mathrm{dl}}$ provides a complementary per-market quantification. The January 2026 U.S.--Venezuela operation cluster, where the U.S.\ Department of Justice indictment of Master Sergeant Gannon Van Dyke \citep{doj2026vandyke} provides a rare external enforcement benchmark on at least one alleged informed trader, illustrates how the layers stack: account-level lifecycle heuristics identify a small set of suspicious accounts with face-valid enforcement alignment; legal investigation establishes whether a specific trader actually possessed non-public information; per-market scoring would quantify how much information was leaked into each contract before public observation. None of the three layers subsumes the others, and a combined surveillance pipeline gains in precision precisely because each layer filters a different dimension of the problem.
\end{abstract}



\section{Introduction}
\label{sec:intro}

Decentralized prediction markets such as Polymarket aggregate dispersed beliefs into continuously updated probability signals on real-world events. They have, since 2024, accumulated a substantial public-reporting record of suspected informed trading: in the hours before the October 2024 Iran strike on Israel, six newly-created wallets purchased YES shares at prices as low as ten cents and realized roughly \$1.2 million when the market resolved \citep{mitts2026iran}. The 2026 U.S.--Iran conflict cluster traded over \$832 million in cumulative volume across markets that public reporting flagged for anomalous pre-event positioning. The January 2026 U.S.--Venezuela operation produced a contemporaneous wave of pre-event purchases on Maduro- and Venezuela-related markets, and culminated, on April 23, 2026, in the first U.S.\ federal indictment for prediction-market insider trading: U.S.\ Army Master Sergeant Gannon Van Dyke was charged with using classified information about \emph{Operation Absolute Resolve} to profit roughly \$409{,}881 across thirteen Polymarket trades \citep{doj2026vandyke, cftc2026vandyke}.

Against this empirical record, the academic literature has, until recently, lacked methodologically distinct detection frameworks. April 2026 saw three approaches surface in close succession.

\subsection{Three approaches surfaced in April 2026}

\citet{mitts2026iran} construct a composite informed-trading screen combining cross-sectional bet size, within-trader bet size, profitability, pre-event timing, and directional concentration. They apply the screen to more than $210{,}000$ wallet--market pairs covering Polymarket activity from 2023 through early 2026 and estimate that approximately \$143 million in aggregate anomalous profit was extracted across the platform. Their methodology operates at the wallet-market pair level, retrospectively, and yields an anomaly score per pair.

\citet{gomezcram2026informed}, in a working paper posted to SSRN on April 20, 2026 and revised on April 25, 2026, develop two related methodologies applied to the complete transaction history of Polymarket, covering 1.72 million accounts, 210{,}322 markets, and \$13.76 billion in trading volume across 2023--2025. Their primary contribution is a sign-randomization skill classifier that re-runs each account's trade history $10{,}000$ times with the buy/sell direction randomized at the event level, and classifies the account as skilled, lucky, unlucky, or unskilled based on the location of realized PnL in the simulated null. They report that $3.14\%$ of accounts ($54{,}477$ in their test sample, restricted to accounts trading in at least ten events) qualify as ``skilled winners'' whose order flow predicts both next-period price changes and final outcomes; together with market makers, this minority captures more than $30\%$ of total platform gains. The classification exhibits striking out-of-sample persistence: $44\%$ of accounts classified as skilled in a training subset retain that classification on a randomly held-out test set. Their secondary contribution, methodologically separate from the skill classifier, is a single-event lifecycle-and-conviction heuristic that flags $1{,}950$ accounts as suspected insiders by independent timing and volume criteria. The two components address structurally different populations and have correspondingly different statistical properties; we discuss them in detail in \Cref{sec:signrand}.

\citet{nechepurenko2026foresightflow_methodology}, in a methodology preprint (April 2026), introduce the \emph{Information Leakage Score} (ILS):
\[
\ILS(M) \;=\; \frac{p(\Tnews) - p(\Topen)}{p_{\Tres} - p(\Topen)},
\]
which quantifies, on a single resolved binary market, the fraction of the terminal information move that was priced in before the corresponding public news event. The authors specify three operational scope conditions for the score, develop a Murphy-decomposition reading that connects ILS to the proper-scoring-rule literature, and extend the score to deadline-resolved contracts via the \emph{deadline-ILS} variant ILS$^{\mathrm{dl}}$, anchored at an article-derived event timestamp. The companion empirical paper \citep{nechepurenko2026foresightflow_empirical} applies the framework end-to-end to the 2026 U.S.--Iran conflict cluster, demonstrating that the article-derived anchor distinguishes signal from proxy artefact on at least one documented case.

\subsection{Shared concern, different layers}

The three approaches address the same substantive concern---the presence of informed trading in a public, on-chain prediction-market venue---but each operates at a different methodological layer.

\citet{mitts2026iran} and the skill classifier of \citet{gomezcram2026informed} are wallet-level methodologies; they produce statements about \emph{who, on the platform, behaved anomalously over their full trading history}. The lifecycle-and-conviction heuristic of \citet{gomezcram2026informed} is also wallet-level but operates at the per-event level rather than the multi-event level; it produces statements about \emph{which accounts exhibited a single-event behavioural signature consistent with informed trading}. The ILS framework \citep{nechepurenko2026foresightflow_methodology} is a market-level methodology; it produces statements about \emph{which markets exhibited information front-loading prior to public observation of the underlying event}.

These layers are not substitutes. Wallet-level skill classification answers ``has this account demonstrated repeated directional ability across many events?''. Wallet-level lifecycle screening answers ``does this account exhibit the behavioural signature of a single-event informed trader?''. Market-level information-leakage scoring answers ``on this specific contract, how much of the eventual move from opening price to resolution had been priced in before public observation?''. The U.S.\ Department of Justice indictment of Van Dyke \citep{doj2026vandyke} illustrates the practical relevance of holding all of these questions simultaneously: prosecution required (i) identifying which specific trader is alleged to have had access to classified information, which is a regulatory and legal investigation rather than a statistical inference; (ii) quantifying which contracts the trader profited on and by how much, which is closer to a per-market measure; and (iii) establishing a behavioural pattern of suspicious activity, which is closer to the lifecycle-and-conviction signature. Each of the academic methodologies under consideration contributes to one or two of these dimensions; none addresses all of them.

\subsection{Contribution}

This paper provides:

\begin{enumerate}[leftmargin=1.4em,itemsep=0.4em]
\item A taxonomy of detection methods for prediction markets that organizes the approaches by methodological layer and unit of analysis (\Cref{sec:taxonomy}).
\item A separate analysis of what the sign-randomization \emph{skill classifier} of \citet{gomezcram2026informed} measures (persistent directional skill conditional on opportunity selection), and what their \emph{single-event lifecycle-and-conviction heuristic} measures (face-valid behavioural screen with unmeasured precision); these are two methodologically distinct components of their paper that should not be conflated (\Cref{sec:signrand}).
\item A category-heterogeneity argument: the Polymarket sample on which all approaches are evaluated is dominated by sports, politics, and crypto markets with structurally different information technologies, so a platform-wide skill classification is mechanism-ambiguous and category-conditioned decompositions are required for surveillance use (\Cref{sec:signrand-skill}).
\item A parallel analysis of the ILS framework (\Cref{sec:ils}).
\item A layered analysis of the January 2026 U.S.--Venezuela operation cluster, where the Van Dyke indictment provides external enforcement evidence on at least one alleged informed trader, the \citet{gomezcram2026informed} heuristic flags three accounts whose realized PnL aligns with the indictment to the dollar on the lead account, and ILS$^{\mathrm{dl}}$ would in principle quantify per-contract front-loading (\Cref{sec:convergence}).
\item A sketch of a combined surveillance pipeline using \emph{category-conditioned account risk scoring} as Stage 1 (rather than a naive ``run sign-randomization classifier and flag heuristic insiders'' workflow), ILS$^{\mathrm{dl}}$ as Stage 2, and human compliance review as Stage 3 (\Cref{sec:pipeline}).
\item Identification of three open questions that none of the methodologies in their current form addresses: resolution-criteria uncertainty, continuous pre-event trade collection, and cross-market coordination (\Cref{sec:open}).
\end{enumerate}

\paragraph{Reproducibility note.} The discussion of \citet{gomezcram2026informed} in this paper is based on the working paper text and tables, including Appendix B (the Maduro case study) and the topic-decomposition tables in the main appendix. The underlying account-level classifications, the labelled list of $1{,}950$ heuristic-flagged accounts, the category-conditioned skill decompositions, and the source code for the sign-randomization procedure have not, to our knowledge, been released as a reproducible artifact. The Polymarket transaction history on which their analysis is built is publicly available on-chain, so the methodology is reproducible \emph{in principle} on an aligned account population by any researcher with sufficient infrastructure for full-history transaction processing; we did not undertake such reproduction in the present comparative work, and we mark this constraint as a binding limitation on operational use of the methodology in \Cref{sec:signrand-reproducibility,sec:pipeline}.

\section{A taxonomy of detection methods}
\label{sec:taxonomy}

We organize prediction-market informed-trading detection methodologies by three structural dimensions: \emph{granularity} (the unit of analysis), \emph{timing applicability} (whether the method operates retrospectively or in real time), and \emph{output type} (anomaly score, skill classification, or interpretable per-unit measure). \Cref{tab:taxonomy} summarizes the three approaches under consideration.

\begin{table}[t]
\centering
\caption{Taxonomy of detection methodologies for informed trading on prediction markets surfaced in 2024--2026. ``Wallet--market'' granularity means the unit of analysis is a (wallet, market) pair; ``account, multi-event'' means the wallet alone, aggregated across markets and conditioned on a minimum-event threshold; ``account, single-event'' means the wallet at the per-event level; ``market'' means the market alone, aggregated across wallets. The two components of \citet{gomezcram2026informed}---the skill classifier and the lifecycle-and-conviction heuristic---are listed separately because they answer different questions and apply to different populations.}
\label{tab:taxonomy}
\small
\renewcommand{\arraystretch}{1.3}
\begin{tabularx}{\linewidth}{@{}p{3.0cm}p{2.6cm}p{2.4cm}p{2.6cm}X@{}}
\toprule
\textbf{Method} & \textbf{Granularity} & \textbf{Timing} & \textbf{Output} & \textbf{Information used} \\
\midrule
Composite anomaly screen \citep{mitts2026iran} & Wallet--market pair & Retrospective & Anomaly score on (wallet, market) pairs & Bet size, profitability, pre-event timing, directional concentration \\
Sign-randomization skill classifier \citep{gomezcram2026informed} & Account, multi-event ($\geq 10$ events) & Retrospective & 4-category skill classification with calibrated $p$-values & Full PnL distribution under randomized event-level directions \\
Lifecycle-and-conviction heuristic \citep{gomezcram2026informed} & Account, single-event & Retrospective & Binary flag: $1{,}950$ accounts in their sample & Account creation time, single-event focus, post-resolution dormancy, volume and profit thresholds \\
ILS / ILS$^{\mathrm{dl}}$ \citep{nechepurenko2026foresightflow_methodology} & Market & Post-event, pre-resolution & Per-market score with scope-condition gate & Price trajectory $[\Topen, \Tres]$, anchor at $\Tnews$ or $\Tevent$ \\
\bottomrule
\end{tabularx}
\end{table}

\subsection{Layers and the questions they answer}

Each row of \Cref{tab:taxonomy} answers a distinct operationally meaningful question about prediction-market informed trading.

\paragraph{Wallet-level retrospective scoring on (wallet, market) pairs} asks: ``Among observed wallet--market pairs, which exhibit a statistical signature of anomalous performance?'' \citet{mitts2026iran}'s composite screen aggregates several features (cross-sectional bet size, within-trader bet size, profitability, pre-event timing, directional concentration) into a single anomaly score and identifies the upper tail. This methodology is well-suited to retrospective surveys of platform-wide insider activity and to estimating economic magnitudes (e.g., the \$143M aggregate figure). It does not provide a real-time signal, because profitability is only known at resolution, and it does not separate skill from luck explicitly.

\paragraph{Wallet-level skill-vs-luck classification on accounts with sufficient trading history} asks: ``Among the population of accounts that have traded enough events to permit statistical inference, what fraction can be distinguished from luck-driven outcomes?'' The sign-randomization classifier of \citet{gomezcram2026informed} addresses this question on accounts trading in at least ten events. The test is statistically well-calibrated, with a clean null hypothesis (event-level randomization of trade direction) and per-account $p$-values, and the authors document a striking skill-persistence finding: $44\%$ of skilled accounts retain their classification out-of-sample, compared to roughly $10\%$ in published studies of equity mutual-fund managers. The methodology is, however, fundamentally retrospective and excludes the one-shot insider population by design.

\paragraph{Wallet-level lifecycle-and-conviction screening at the per-event level} asks: ``Among accounts trading a specific event, which exhibit the behavioural signature (recent creation, single-event focus, post-resolution dormancy, sufficient volume and conviction) of a one-shot informed trader?'' The lifecycle heuristic in \citet{gomezcram2026informed} addresses this question. It is methodologically separate from the skill classifier: it applies to accounts the skill classifier excludes, uses different inputs (account creation timestamps and per-event activity windows rather than full PnL distributions), and produces a binary flag rather than a graded skill score. Its precision against an external ground-truth set has not been reported in the released paper.

\paragraph{Market-level information front-loading} asks: ``On a specific resolved market, what fraction of the terminal information move was priced in before the corresponding public news event?'' The ILS framework \citep{nechepurenko2026foresightflow_methodology} operates at this layer. The output is a per-market score with explicit scope conditions; it is interpretable in terms of the Murphy decomposition of the Brier score; and the relevant price trajectory is observable once $\Tevent$ has been identified, enabling post-event, pre-resolution scoring without requiring final resolution outcomes. Wallet-level attribution is not part of the methodology proper.

\subsection{What the layers leave to each other}

Each layer produces information the other two cannot produce, and each leaves to the others questions it does not address.

The wallet-level approaches do not, on their own, identify \emph{which contracts} the wallets traded with informational advantage; they aggregate across all markets in which a wallet was active. A skilled wallet whose skill consists of identifying mispricings in NBA prop markets is, by sign-randomization, indistinguishable from a wallet whose skill consists of trading on classified geopolitical information. The market-level approach makes the question per-market explicit: a high ILS$^{\mathrm{dl}}$ on a specific Maduro market is a different statement from a high ILS$^{\mathrm{dl}}$ on a NBA market.

The market-level approach, conversely, does not on its own identify \emph{which wallets} were responsible for the front-loading. A market with high ILS$^{\mathrm{dl}}$ may have been moved by a single wallet with classified information, by a coordinating group of wallets, or by a large number of independent traders responding to a leak in semi-private channels. Wallet-level methods address this in the same way the present comparison treats their methodology: as a complementary layer, not a substitute.

\section{What sign-randomization measures}
\label{sec:signrand}

We summarize the methodology of \citet{gomezcram2026informed} as documented in their working paper, then characterize what the methodology captures well and what it cannot do. The discussion in this section relies on the working paper text and on the appendix tables therein; we have not independently reproduced the underlying classification on Polymarket data.

The Gomez-Cram et al.\ paper combines two methodologically distinct components: a sign-randomization \emph{skill classifier} that operates over each account's full trading history, and a separate \emph{single-event lifecycle-and-conviction heuristic} that flags accounts the skill classifier explicitly cannot identify. We discuss the two components separately because they answer different questions, have different statistical properties, and require different operational treatment.

\subsection{Sign-randomization skill classifier}
\label{sec:signrand-skill}

For each account in the Polymarket population, the authors construct the realized profit-and-loss trajectory across the account's full trading history. They then construct a counterfactual distribution of PnL trajectories by holding all trades fixed in event, market, contract size, price, and timing, and randomizing only the buy/sell direction. The randomization is applied at the \emph{event level}: all trades within a single event are correlated under the null, both to capture position-building within a market and to capture related bets across multiple markets within the same event (e.g., simultaneously buying ``Trump wins'' and ``Harris loses''). The procedure is repeated $10{,}000$ times.

The realized PnL is then compared to the simulated null distribution and the account is assigned a $p$-value. Accounts with $p < 0.05$ are classified as \emph{skilled winners} ($54{,}477$ accounts in the test sample, $3.14\%$ of the population); accounts with $p > 0.95$ are \emph{unskilled losers} ($110{,}703$ accounts, $6.4\%$); the remainder are \emph{lucky winners} (positive realized PnL, no statistical significance, $29.0\%$) or \emph{unlucky losers} (negative realized PnL, no significance, $61.4\%$). A residual category of market makers ($0.1\%$) is identified separately by liquidity-provision behaviour.

The headline aggregate findings of the classifier are: skilled winners plus market makers, comprising under $3.5\%$ of accounts, capture more than $30\%$ of platform gains; lucky winners ($29\%$ of accounts) capture the remaining $69\%$; unlucky and unskilled losers ($67\%$ of accounts) absorb the entirety of platform aggregate losses. The skill classification exhibits out-of-sample persistence: of accounts classified as skilled in a training subset of events, $44\%$ retain that classification on a randomly held-out test set, in contrast to roughly $10\%$ retention for skilled mutual-fund managers in published equity-fund persistence studies.

\paragraph{What the skill classifier measures.}
The natural reading of the skill test is that it measures \emph{persistent directional profitability conditional on the account's chosen opportunity set}. The randomization holds fixed which events the account chose to trade, which markets within those events, what size positions it took, and when it took them; it varies only the directional side (buy vs.\ sell) of each event-level decision. The test therefore answers the question: ``given this account's chosen events, markets, sizes, and timing, was its realized direction unusually profitable relative to a randomized-direction counterfactual?''

This is methodologically stronger than naive top-PnL ranking---which treats winners and losers symmetrically without correcting for the bimodal resolution structure of binary outcome markets, and which would incorrectly attribute large lucky payouts to skill---but it is also \emph{narrower} than a direct test of insider trading. The test does not ask, and does not answer:

\begin{itemize}[leftmargin=1.4em,itemsep=0.3em]
\item whether the trader's information was private rather than public;
\item whether private information, if held, was illegally obtained or constituted material non-public information (MNPI) in the regulatory sense;
\item whether the trader was an insider in any specific operational definition (e.g., the U.S.\ statutory ``Eddie Murphy Rule'' under Dodd-Frank Section $746$);
\item on which specific markets within an account's history the directional skill was concentrated;
\item whether the directional skill reflects superior public-information processing (e.g., faster news ingestion, better statistical modelling, on-chain analytics for crypto markets, public weather-model output for meteorological markets) or genuinely non-public information.
\end{itemize}

A skilled sports bettor specialised in NBA injury-news ingestion, a crypto on-chain analyst trading whale-flow patterns, a weather modeller trading hurricane probabilities, an arbitrageur exploiting cross-platform pricing inefficiencies, a journalist with industry contacts, and a classified-information geopolitical insider can all appear as ``skilled winners'' under sign-randomization. The mechanisms behind their skill are categorically different, and the regulatory significance of the trading is correspondingly different.

\paragraph{Category heterogeneity in the underlying sample.}
The Polymarket sample on which the classifier is evaluated is structurally heterogeneous in its information content. Sports markets account for $42.05\%$ of all markets and $35.37\%$ of total trading volume; politics markets account for $7.52\%$ of markets and $35.85\%$ of volume; crypto markets account for $36.96\%$ of markets and $18.73\%$ of volume; the remaining categories (culture, economy, technology, finance) jointly account for under $10\%$ of volume \citep[Table A.2]{gomezcram2026informed}. Together, sports, politics, and crypto cover roughly $89\%$ of platform volume, and these categories rest on entirely different information technologies. ``Skill'' in NBA prop markets is the ability to ingest injury news and lineup updates faster than the consensus; ``skill'' in cryptocurrency markets is the ability to read on-chain analytics and exchange flow signals; ``skill'' in geopolitical markets may, on the upper tail, involve access to genuinely non-public state information.

A platform-wide skilled-winner classification therefore conflates qualitatively different populations. The $54{,}477$ skilled accounts include sports bettors, crypto analysts, weather modellers, scheduled-news traders, arbitrageurs, and on the upper tail, accounts whose realized profit is consistent with informed trading on private information. Without category-conditioned decomposition---a per-category classification that reports skilled-winner rates, profit shares, and persistence rates separately for sports, politics, crypto, and the remaining categories---the platform-wide statistic is mechanism-ambiguous and cannot be used as a direct surveillance signal.

\paragraph{One-shot insiders are excluded from the skill test by construction.}
A statistically reliable sign-randomization $p$-value requires repeated decisions: the test gains power as the number of independent event-level decisions in an account's history increases. \citet{gomezcram2026informed} restrict their main classification analysis (Figure $2$ of their paper) to accounts that traded in at least ten events, on the grounds that fewer events do not permit reliable skill-vs-luck separation. This is statistically reasonable; it has, however, an immediate operational consequence for surveillance: a one-shot informed trader---an account opened specifically to trade on classified information for a single event, then closed---fails the ten-event threshold and falls outside the skill classifier's scope. The authors themselves note in their Maduro case-study analysis that the three suspicious accounts are classified as \emph{lucky winners} rather than \emph{skilled winners}, ``as the sign-randomization test cannot identify skill from a single informed trade'' \citep[Section 4.4]{gomezcram2026informed}.

This is not a defect of sign-randomization; it is a design choice with a known scope. But it means that the skill classifier alone is unsuited to detecting the canonical insider-trading pattern that has dominated the public-reporting record on Polymarket: an account opened in the days before a single high-stakes event, taking a concentrated position, and going dormant after resolution.

\subsection{Single-event lifecycle-and-conviction heuristic}
\label{sec:signrand-heuristic}

To address the population the skill classifier cannot reach, \citet{gomezcram2026informed} apply a separate two-criterion heuristic. An account is flagged as a ``suspected insider'' if it satisfies both: \emph{timing}---the account is created within seven days before an event and places no further trades after that event resolves---and \emph{conviction}---the account trades exclusively in that one event, with at least \$$1{,}000$ in volume and at least \$$1{,}000$ in realized profit. The authors report that this heuristic flags $1{,}950$ accounts across the Polymarket sample, with mean realized profit per flagged account of \$$15{,}012.92$ and median of \$$2{,}758.29$. Insider order imbalance (i.e., the net buy direction of flagged accounts) predicts both next-period price changes ($t = 2.54$) and final market outcomes ($t = 8.65$) at high statistical significance; insider trading volume does not significantly reduce absolute pricing error.

\paragraph{What the heuristic measures.}
The heuristic identifies accounts whose \emph{lifecycle behaviour} is consistent with a stylized informed trader: open shortly before an event, trade concentrated in that one event, exit after resolution. It is a behavioural signature, not a mechanism test. The authors are explicit that this is a face-valid screen rather than a calibrated insider classifier, and they motivate it by reference to the Maduro case study before applying it to the full sample.

\paragraph{What the heuristic does not measure.}
The lifecycle-and-conviction signature is consistent with several populations besides classified-information insiders:

\begin{itemize}[leftmargin=1.4em,itemsep=0.3em]
\item \emph{Curious new users} who join the platform during a viral or high-attention event, place a single sizable bet, and do not return when the event resolves. The 2024 U.S.\ presidential cycle and the 2026 Iran conflict cluster both attracted substantial influxes of first-time accounts.
\item \emph{Sybil and promotional farms}: accounts opened structurally on a per-event basis to harvest airdrop campaigns, referral bonuses, or platform promotions, which by design exhibit single-event activity and post-resolution dormancy.
\item \emph{Single-strategy bots} deployed for one specific event and then disabled.
\item \emph{Privacy-rotating accounts} of legitimate traders who, for operational-security reasons, segregate their activity into per-event accounts.
\item \emph{Single-event public-news traders} who placed an informed bet based on publicly available information that the consensus had not yet absorbed---a population that overlaps in behaviour with insiders but is not regulatorily equivalent.
\end{itemize}

The aggregate $1{,}950$ count is therefore not a precise estimate of insider accounts; it is an upper bound on the population of accounts \emph{whose lifecycle behaviour is consistent with informed single-event trading}. The strong predictive performance of insider order imbalance on price and outcome (reported in the authors' Table 6, Panel B) is consistent with the heuristic having high recall on real informed trades---true insiders, when present, do exhibit this lifecycle pattern---but the precision of the heuristic against an external ground-truth set has not, to our knowledge, been reported. Without such a precision study, the operational use of the $1{,}950$ accounts as a surveillance flag would yield an unknown but plausibly substantial false-positive rate.

\subsection{What sign-randomization, as a whole, captures and does not capture}

Combining the two components, the \citet{gomezcram2026informed} methodology contributes the following:

\begin{itemize}[leftmargin=1.4em,itemsep=0.3em]
\item A statistically calibrated population-level distribution of persistent directional skill, with strong evidence that price discovery on Polymarket is concentrated in a small minority and that the platform aggregate cannot be modelled as a uniform-skill / lucky-draw process.
\item A documented out-of-sample skill-persistence finding ($44\%$ retention) that contrasts sharply with the corresponding mutual-fund literature.
\item A face-valid lifecycle heuristic that captures conduct consistent with informed trading when applied to the Maduro case study, and whose order-imbalance regressions show strong contemporaneous price-prediction power.
\end{itemize}

The methodology does not, in either component, provide:

\begin{itemize}[leftmargin=1.4em,itemsep=0.3em]
\item A direct identification of insider trading in the regulatory sense; only conditional directional skill (skill classifier) or lifecycle suspicion (heuristic).
\item A category-conditioned decomposition; the platform-wide skilled-winner statistic is mechanism-ambiguous in a sample dominated by sports, politics, and crypto.
\item A real-time signal on active markets; both components require resolved outcomes and final PnL.
\item A per-market quantification; the unit of analysis is the account, aggregated across markets.
\item A separation of resolution-criteria uncertainty from informed front-loading; markets that misprice for criteria-ambiguity reasons (Iran-Apr30 in \citet{nechepurenko2026foresightflow_empirical} is an example) generate spurious skill or unskill classifications under the test.
\end{itemize}

\subsection{Reproducibility}
\label{sec:signrand-reproducibility}

\citet{gomezcram2026informed} describe their data construction and methodology at the level of detail customary for a finance working paper. The underlying account-level skill classifications, the labelled list of $1{,}950$ heuristic-insider accounts, the category-conditioned decompositions, and the source code for the sign-randomization procedure have not, to our knowledge, been released as a reproducible artifact. The Polymarket transaction history on which their analysis is built is publicly available on-chain, so the methodology is reproducible \emph{in principle} on an aligned account population by any researcher with sufficient infrastructure for full-history transaction processing; we did not undertake such reproduction in the present work.

For operational use as a surveillance layer, the absence of released account labels is a binding constraint: the $1{,}950$ flagged accounts cannot be cross-referenced against an external compliance database or against subsequent enforcement actions without independent reproduction. The strongest research finding of \citet{gomezcram2026informed}---that price discovery on Polymarket is concentrated in a small minority of skilled accounts---stands on its own as a contribution to the platform-economics literature regardless of whether the underlying classifications are released. The operational use of the methodology in a surveillance pipeline, however, requires either the release of labels or independent reproduction with category-conditioned decomposition. We return to this point in \Cref{sec:pipeline}.

\section{What ILS$^{\mathrm{dl}}$ measures}
\label{sec:ils}

We now perform the parallel analysis on the deadline-ILS framework of \citet{nechepurenko2026foresightflow_methodology, nechepurenko2026foresightflow_empirical}.

\subsection{Methodology recap}

For a deadline-resolved binary market $M$ with first trade at $\Topen$, contractual deadline $\Tdead$, recovered public-event timestamp $\Tevent \in [\Topen, \Tdead]$, formal resolution at $\Tres$, and binary outcome $p_{\Tres} \in \{0, 1\}$:
\[
\ILS^{\mathrm{dl}}(M) \;=\; \frac{p(\Tevent^-) - p(\Topen)}{p_{\Tres} - p(\Topen)},
\]
where $p(\Tevent^-)$ is the market price one minute before public observation of the underlying event. The score is admitted only when three scope conditions are all met: (i) the \emph{edge-effect} condition $|p(\Topen) - 0.5| \leq 0.4$, requiring substantive uncertainty at market opening; (ii) the \emph{non-trivial-move} condition $|\Delta_{\mathrm{total}}| \geq \varepsilon$ (we use $\varepsilon = 0.05$), requiring a measurable terminal move so that the denominator of the score is well-conditioned; and (iii) the \emph{anchor-sensitivity} condition that the score be robust to small perturbations of $\Tevent$. Markets that fail any of these conditions---including markets with $|\Delta_{\mathrm{total}}| < \varepsilon$ (trivial total move), opening prices near $0$ or $1$ (consensus markets), or numerically unstable scores under anchor perturbation---are excluded from analysis rather than scored unreliably. The framework adopts $\theta_{\Topen} \equiv p(\Topen)$ as a conservative baseline and a per-category exponential survival function $S(\tau) = \exp(-\lambda \tau)$ for the time-to-event distribution, with hazard rate $\lambda$ fitted by maximum likelihood on a comparable population.

The empirical companion paper \citep{nechepurenko2026foresightflow_empirical} demonstrates the framework end-to-end on the largest applicable case in the FFIC inventory: ``US forces enter Iran by April 30,'' a \$269M-volume contract for which $\Tevent = $ April 3, 2026 (F-15E special operations entry) was recovered via LLM-assisted multi-source verification. The article-derived anchor yields $\ILS^{\mathrm{dl}} = +0.113$; the resolution-anchored proxy ($\Tres - 1\,\mathrm{h}$) yields $-0.331$; the two values lie on opposite sides of zero and differ by $0.444$ in magnitude.

\subsection{What the methodology captures well}

\paragraph{Per-market interpretability.}
The output of the framework is one number per market, with an explicit interpretation: $\ILS^{\mathrm{dl}}(M) = 0.6$ means roughly $60\%$ of the eventual move from opening price to resolution had been priced in before public observation of the underlying event. This is the natural unit for a regulator or platform officer asking a contract-specific question, and it is the natural unit for a research investigator constructing case studies.

\paragraph{Post-event, pre-resolution applicability.}
The score depends only on the price trajectory up to time $t$ and on the recovered $\Tevent$ when known; it does not require resolution outcome or final PnL. For a market that is currently active, $\ILS^{\mathrm{dl}}$ can be computed up to $\Tevent$ as soon as the public event is identified. We emphasize that this is \emph{post-event, pre-resolution} applicability rather than pre-event real-time monitoring: the score is well-defined after the underlying event has been identified in public reporting, but before the contract has formally resolved through the platform's oracle. A pre-event alarm in the strict sense---a signal raised \emph{before} the underlying event occurs, based on price-trajectory features alone---is a separate methodological problem that the framework does not currently address. The companion methodology paper develops the architectural consequences of post-event causal inference: every feature used at inference time is causal in the public information available at that time.

\paragraph{Scope conditions are explicit, not implicit.}
The three scope conditions (edge effect, trivial resolution, anchor sensitivity) are stated as preconditions for interpreting the score; markets failing them are excluded from analysis rather than scored unreliably. This is methodologically conservative: the framework prefers a smaller domain of confident application to a larger domain of noisy results.

\paragraph{Connection to proper scoring rules.}
Through the Murphy decomposition of the Brier score, ILS admits an interpretation as the share of the resolution component of the Brier score accumulated before $\Tnews$ \citep{nechepurenko2026foresightflow_methodology}. This connects label generation to a well-developed literature on probabilistic forecasting and provides a theoretical grounding the score does not borrow from elsewhere.

\paragraph{Resolution-criteria uncertainty is separable.}
The Iran-Apr30 case study \citep{nechepurenko2026foresightflow_empirical} demonstrates that ILS$^{\mathrm{dl}}$ measures something distinct from market-level resolution-criteria uncertainty. A market that mispredicts its eventual outcome because participants are uncertain about resolution criteria can still produce a coherent $\ILS^{\mathrm{dl}}$ at the article-derived anchor; the score does not conflate the two phenomena.

\subsection{What the methodology cannot do}

\paragraph{Population-level skill distribution is not produced.}
The score does not, on its own, classify wallets or characterize the dispersion of trader skill on the platform. For population-economics questions of the form ``what fraction of accounts is skilled?'', ILS$^{\mathrm{dl}}$ is not the relevant tool.

\paragraph{$\Tevent$ measurement error.}
The framework anchors the score at an article-derived $\Tevent$ recovered from public reporting. In practice, public information about an event diffuses through multiple channels that may precede a publishable news article: real-time social media (Telegram channels, Twitter, Discord), government press hints, on-the-record briefings to wire services, and informed participant commentary in the market itself. The recovered $\Tevent$ is therefore the timestamp of the earliest reproducible news source, not necessarily the timestamp at which the information first became available to a sufficiently networked participant. To the extent that information leaks earlier through informal channels, the article-derived anchor systematically over-estimates the pre-event window and under-estimates the leakage measured by ILS$^{\mathrm{dl}}$. The companion empirical paper notes this concern but does not characterize its magnitude; an explicit study of inter-source timestamp dispersion would be a natural next-stage refinement.

\paragraph{Denominator instability.}
The scope condition $|\Delta_{\mathrm{total}}| \geq \varepsilon$ guards the denominator of the score against numerical instability when the terminal move is small. The threshold $\varepsilon = 0.05$ is a practical choice and not a calibrated cut. Markets near this threshold produce ILS$^{\mathrm{dl}}$ values that are admitted under the condition but are sensitive to small perturbations in $p(\Topen)$ or $p_{\Tres}$; the companion methodology paper recommends reporting the score together with a sensitivity range over the threshold.

\paragraph{Interpretation of negative and over-unity values.}
The score can take values in $(-\infty, +\infty)$ in principle. Negative values indicate that $\Delta_{\mathrm{pre}}$ moved in the opposite direction from $\Delta_{\mathrm{total}}$: the price was moving against the eventual outcome before the public event. Values greater than one indicate that the price had already moved beyond the resolution level before the public event and subsequently retraced. Both regimes are diagnostic but require careful interpretation. In particular, ILS$^{\mathrm{dl}} > 1$ is consistent with either strong informed front-loading followed by partial retracement, or with resolution-criteria uncertainty causing the market to overshoot before re-anchoring on the final outcome \citep[the Iran-Apr30 case study in][illustrates the latter]{nechepurenko2026foresightflow_empirical}. The methodology paper provides the formal interpretation of these regimes but does not, in the present companion empirical paper, examine the populations of negative-ILS or over-unity-ILS markets in detail.

\paragraph{Wallet attribution requires per-trade data not currently in scope.}
The companion empirical paper \citep{nechepurenko2026foresightflow_empirical} reports that wallet-level features for the FFIC cluster were constrained by the available trade history to the resolution-settlement window, owing to The Graph subgraph indexer's coverage policy. The signature ``cross-market wallet coordination at $\Tevent^-$'' that one would naturally want to test alongside ILS$^{\mathrm{dl}}$ is not retrievable retrospectively from the Polymarket subgraph; it requires continuous per-trade collection from $\Topen$ onward as a separable infrastructure investment.

\paragraph{Sample-size limited by pipeline requirements.}
Computing ILS$^{\mathrm{dl}}$ end-to-end requires recovered article-derived $\Tevent$, complete CLOB price coverage from $\Topen$, scope-condition compliance, and positive lead time $\tau = \Tevent - \Topen$. Of the eighteen substantive markets in the 2026 U.S.--Iran cluster at which the empirical companion paper attempts $\Tevent$ recovery, only one satisfies all four requirements. The pipeline is methodologically demanding, and population-scale empirical evaluation of the framework is gated on infrastructure improvements rather than on the methodology itself.

\paragraph{Article-derived $\Tevent$ recovery is not free.}
Each recovered $\Tevent$ involves an LLM call with multi-source web verification (approximately \$$0.09$ per market in the companion paper's pipeline) and human review for high-stakes cases. This compares unfavourably to sign-randomization, which requires only the on-chain trade history that is already public.

\subsection{Where these limits matter}

For population-level economics questions about prediction-market structure, the framework is silent. For per-account skill ranking across the full corpus, the framework is silent. The methodology operates one market at a time, and for many empirical questions of platform interest, that is the wrong unit.

\section{The Maduro / Venezuela Cluster as a Layered Test Case}
\label{sec:convergence}

The January 2026 U.S.--Venezuela operation cluster offers a case study where the methodological layers under consideration can be observed to stack constructively, and where the U.S.\ Department of Justice indictment of Master Sergeant Gannon Van Dyke \citep{doj2026vandyke} provides an external enforcement benchmark on at least one alleged informed trader. We examine the cluster from each layer in turn. We emphasize that the cluster is a layered \emph{test case} rather than a completed three-method convergence: the lifecycle-and-conviction screen of \citet{gomezcram2026informed} and the legal fact pattern of the DOJ indictment are observed in the cluster, while ILS$^{\mathrm{dl}}$ is only \emph{proposed} on these markets and not actually computed in the present comparative paper.

\subsection{The cluster as a cross-methodological test case}

\emph{Operation Absolute Resolve} was a U.S.\ military operation, executed in the predawn hours of January 3, 2026, that resulted in the apprehension of Venezuelan President Nicolás Maduro and his wife in Caracas. The operation had been in classified planning since at least December 8, 2025, with active execution-planning involvement of multiple U.S.\ Army Special Forces personnel. On April 23, 2026, the U.S.\ Department of Justice unsealed an indictment charging Master Sergeant Gannon Van Dyke, an active-duty Army Special Forces soldier with documented operational involvement in the planning and execution of \emph{Operation Absolute Resolve}, with thirteen Polymarket trades placed between December 27, 2025 and January 2, 2026. Van Dyke realized approximately \$$409{,}881$ in profit on these trades. The markets named in the DOJ indictment include ``U.S.\ Forces in Venezuela by January 31, 2026,'' ``Maduro out by January 31, 2026,'' ``Trump invokes War Powers against Venezuela by\ldots,'' and ``Will the U.S.\ invade Venezuela by January 31, 2026.''

The corresponding Polymarket contracts in the FFIC inventory \citep{nechepurenko2026foresightflow_methodology} are $11$ markets with approximately \$$89$M cumulative volume: $7$ contracts that resolved YES on or shortly after January 3, 2026, and $4$ context contracts (deadline-miss or strict-criterion NO outcomes) that resolved NO. The cluster is thus a deadline-contract cluster of the form documented as the dominant structural type in the public-reporting record on Polymarket insider trading.

The convergence between the three methodological approaches is summarized in \Cref{tab:maduro-convergence}. We treat each in turn.

\begin{table}[t]
\centering
\caption{Layered methodological readings of the January 2026 U.S.--Venezuela operation cluster. The DOJ indictment of Van Dyke (April 23, 2026) provides external enforcement evidence on at least one alleged informed trader. Numerical claims about \citet{gomezcram2026informed} are drawn from the working paper text and tables, including Appendix B; account-level labels for the $1{,}950$ flagged accounts are not part of the released paper.}
\label{tab:maduro-convergence}
\small
\renewcommand{\arraystretch}{1.3}
\begin{tabularx}{\linewidth}{@{}p{4cm}p{4.6cm}X@{}}
\toprule
\textbf{Approach} & \textbf{Output on the cluster} & \textbf{What it does not produce} \\
\midrule
Sign-randomization skill classifier \citep{gomezcram2026informed} & Skill classifier excludes one-shot accounts (ten-event minimum); the three Maduro-cluster accounts therefore appear as ``lucky winners'' in the skill classification, not skilled & Detection of one-shot informed trading; per-contract front-loading; whether \emph{specific} markets were anomalously priced versus reasonable consensus \\
Lifecycle-and-conviction heuristic \citep{gomezcram2026informed} & Three accounts flagged: \texttt{0x31a...b8ed9}, \texttt{0x6ba...a94c5}, \texttt{0xa72...febd4}. Combined PnL \$$626{,}484$. The first account's PnL of \$$409{,}882$ matches the DOJ indictment's quoted Van Dyke profit of \$$409{,}881$ to the dollar. & Per-contract front-loading; precision against full $1{,}950$-account flagged population (Maduro provides face validity, not full-population precision) \\
DOJ indictment \citep{doj2026vandyke} & One trader identified by name and role; \$$409{,}881$ profit across 13 trades; specific markets named & A general detection methodology; the indictment provides external enforcement evidence on this case but not a generalizable test \\
ILS$^{\mathrm{dl}}$ framework \citep{nechepurenko2026foresightflow_methodology, nechepurenko2026foresightflow_empirical} & Would provide per-market $\ILS^{\mathrm{dl}}$ on the $7$ YES-resolved contracts; not executed in this paper & Wallet-level identification of which accounts moved each market \\
\bottomrule
\end{tabularx}
\end{table}

\subsection{What sign-randomization identifies on the cluster}

\citet[Appendix B and Table B.1]{gomezcram2026informed} report a case study of three Polymarket accounts that took sizable positions in Maduro-related markets in the days and hours before the January 3, 2026 operation. The three accounts and their reported activity, as documented in the working paper, are summarized in \Cref{tab:maduro-three-accounts}.

\begin{table}[t]
\centering
\caption{The three accounts flagged by \citet[Appendix B, Table B.1]{gomezcram2026informed} in the Maduro / Venezuela cluster case study. All three traded the contract ``Maduro out by January 31, 2026?'' before the public announcement of \emph{Operation Absolute Resolve} on January 3, 2026. Account addresses are reported in abbreviated form in the working paper; the full addresses have not been released.}
\label{tab:maduro-three-accounts}
\small
\renewcommand{\arraystretch}{1.3}
\begin{tabularx}{\linewidth}{@{}llXrr@{}}
\toprule
\textbf{Address (abbr.)} & \textbf{Created (UTC)} & \textbf{Last traded (UTC)} & \textbf{PnL (\$)} & \textbf{Volume (\$)} \\
\midrule
\texttt{0x31a...b8ed9} & 2025-12-27 05:05 & 2026-01-04 00:03 & $409{,}882.03$ & $33{,}933.26$ \\
\texttt{0x6ba...a94c5} & 2026-01-03 03:42 & 2026-01-12 22:30 & $141{,}619.92$ & $25{,}089.86$ \\
\texttt{0xa72...febd4} & 2026-01-02 21:44 & 2026-01-02 21:44 &  $74{,}982.34$ &  $5{,}782.66$ \\
\midrule
\multicolumn{3}{l}{Total profit across the three accounts}        & \multicolumn{1}{r}{$626{,}484.29$} & \\
\bottomrule
\end{tabularx}
\end{table}

All three accounts satisfy the timing-and-conviction heuristic of \Cref{sec:signrand-heuristic}: each was created within seven days before the Maduro event; each traded predominantly in the ``Maduro out by January 31, 2026?'' contract; each met the volume and profit thresholds; and two of the three went dormant entirely after their Maduro positions resolved \citep[the third placed a single \$$4{,}000$ trade in an unrelated market before going dormant; see][Section B]{gomezcram2026informed}. The accounts' realized profits collectively exceed \$$630{,}000$, with conviction visible in the working paper's Figure B.1: the three accounts placed an unusually large order of approximately $80{,}000$ shares late on January $1$, followed by six transactions of comparable size around midnight on January $3$, before the price rose sharply.

\Cref{tab:evidentiary-status} summarizes the evidentiary status of each component of the layered analysis on this cluster.

\begin{table}[t]
\centering
\caption{Evidentiary status of each layered component on the Maduro / Venezuela cluster. ``Reported'' means the claim is documented in the cited source; ``not released'' means the underlying labels or data are not publicly available; ``not executed'' means the analysis is proposed in the present comparative paper but not carried out.}
\label{tab:evidentiary-status}
\small
\renewcommand{\arraystretch}{1.3}
\begin{tabularx}{\linewidth}{@{}p{4.7cm}p{4.0cm}X@{}}
\toprule
\textbf{Component} & \textbf{Status} & \textbf{Source / note} \\
\midrule
Sign-randomization skill classification of cluster accounts & Reported (lucky winners, not skilled) & \citet[Section 4.4]{gomezcram2026informed}; one-shot accounts excluded from skill test by ten-event threshold \\
Lifecycle-and-conviction heuristic case study (3 accounts) & Reported with full Appendix B detail & \citet[Appendix B, Table B.1]{gomezcram2026informed} \\
Heuristic-flagged account labels (full $1{,}950$ population) & Not released & Account labels, code, and category-conditioned decompositions are not part of the released paper \\
Maduro account-to-trader identity mapping & Not released; consistent with public-record evidence to the dollar on lead account & \citet{gomezcram2026informed}; \citet{doj2026vandyke} \\
DOJ enforcement on Van Dyke (alleged informed trader) & Public allegation, not yet conviction & \citet{doj2026vandyke}; \citet{cftc2026vandyke} \\
ILS$^{\mathrm{dl}}$ scores on the seven YES-resolved cluster contracts & Not executed & Proposed in \Cref{sec:convergence-ils} of the present paper; pipeline expansion required \\
\bottomrule
\end{tabularx}
\end{table}

\paragraph{Face validity but no exact-identity attribution from the paper alone.}
The temporal alignment between the first flagged account and the U.S.\ Department of Justice indictment is striking. The first row of \Cref{tab:maduro-three-accounts} reports a realized PnL of \$$409{,}882.03$; the DOJ indictment of Master Sergeant Gannon Van Dyke alleges that he ``profited approximately \$$409{,}881$'' across thirteen Polymarket trades on Maduro- and Venezuela-related contracts \citep{doj2026vandyke}. The two figures match to the dollar, and the dates are consistent with Van Dyke's alleged trading window of December 27, 2025 through January 2, 2026 \citep{cftc2026vandyke}. Taken together, the public-record evidence is consistent with the first flagged account corresponding to an account allegedly used by Van Dyke. We do not, however, have access to a verified mapping between the abbreviated address \texttt{0x31a...b8ed9} and the specific account(s) named in the legal proceedings, and \citet{gomezcram2026informed} do not in their working paper assert such a mapping; we therefore restrict ourselves to noting the consistency of the public-record evidence rather than asserting an identity match.

\paragraph{Heuristic recall and what the case study does and does not demonstrate.}
The Maduro case study provides strong face-valid evidence that the lifecycle-and-conviction heuristic captures conduct consistent with informed trading: at least one of the three flagged accounts is highly likely, on the public-record evidence, to correspond to a trader subsequently charged criminally in connection with classified information about the underlying event. This is informative about the heuristic's \emph{recall} on enforcement-aligned cases---when a documented insider trades, the heuristic appears to flag the relevant accounts---but it does not establish the heuristic's \emph{precision} on the full $1{,}950$-account flagged population. Most of the flagged accounts will not have associated DOJ enforcement actions; whether the remaining $\sim 1{,}947$ accounts include other genuine insiders, sybil-farm accounts, viral-event newcomers, or single-strategy bots cannot be determined from the case study alone.

\paragraph{What sign-randomization does not produce on the cluster.}
The sign-randomization skill classifier proper does \emph{not} identify the Maduro accounts; the three accounts trade in only one event each and therefore fail the ten-event threshold for the skill test. The authors explicitly note this: ``Under our baseline classification, these accounts appear in the lucky winners group rather than among skilled winners, as the sign-randomization test cannot identify skill from a single informed trade'' \citep[Section 4.4]{gomezcram2026informed}. The Maduro detection is therefore wholly attributable to the lifecycle-and-conviction heuristic, not to the skill classifier. This reinforces the methodological-separation point of \Cref{sec:signrand}: the two components of \citet{gomezcram2026informed} answer different questions and apply to different populations.

The Gomez-Cram analysis, taken together, also does not produce a per-contract decomposition of the cluster. Van Dyke's alleged trades span $4$ distinct contracts with different deadlines, different specific event criteria (Maduro custody vs.\ U.S.\ forces in Venezuela vs.\ U.S.\ invasion vs.\ presidential War Powers invocation), and different YES/NO outcomes. From the case study alone, one cannot say ``on this specific deadline contract, $X\%$ of the move from opening price to resolution had been priced in before January 3, 2026 at 10:00 UTC, while on this other contract the corresponding figure is $Y\%$.''

\subsection{What ILS$^{\mathrm{dl}}$ would provide on the cluster}
\label{sec:convergence-ils}

We propose, but do not execute in this paper, the application of the ILS$^{\mathrm{dl}}$ framework to the $7$ YES-resolved markets in the cluster. The pipeline is identical to the Iran-Apr30 application reported in \citet{nechepurenko2026foresightflow_empirical}: for each market, recover the article-derived $\Tevent$ (in the Maduro cluster, $\Tevent$ is January 3, 2026 at approximately 10:00 UTC, the time of the public announcement), check scope conditions, compute $\ILS^{\mathrm{dl}}$ against the article-derived anchor, and report the per-market score together with short-window variants.

The expected output, by methodological design, is a vector of seven per-contract scores quantifying, individually for each contract in the cluster, how much of the eventual move from opening price to resolution had been priced in before January 3, 2026 at 10:00 UTC. This per-contract decomposition is structurally absent from both components of \citet{gomezcram2026informed} and structurally absent from the DOJ indictment's per-trader summary.

We defer the actual computation to a separate work because the seven-market application requires (i) Tier 3 LLM-assisted $\Tevent$ verification on each market, (ii) confirmation of CLOB price coverage from $\Topen$ for each market, (iii) scope-condition checks per market, and (iv) a methodologically careful accounting of the resolution-criteria differences across the seven contracts. The companion empirical paper's reported coverage rate of $18\%$ of positive-$\tau$ markets having full CLOB price coverage suggests that not all seven markets will be evaluable end-to-end with the present infrastructure; partial application is the realistic expectation.

\subsection{How the layers stack on this cluster}

The DOJ indictment of Van Dyke alleges, on the basis of a fact-finding investigation, that a specific named individual had access to classified information and traded on it. The lifecycle heuristic of \citet{gomezcram2026informed} flags a small set of three accounts behaviourally consistent with informed trading, the leading account's PnL aligning with the indicted profit figure to the dollar. The ILS$^{\mathrm{dl}}$ methodology, in the application proposed but not executed in this paper, would identify which contracts in the cluster moved most strongly relative to public information arrival, and would quantify the per-contract magnitude.

These layers are in principle complementary; on the present cluster only two of them are empirically observed (lifecycle screening and legal investigation). The role of ILS$^{\mathrm{dl}}$ in this case is to specify the missing per-contract quantification layer, not to claim it has been measured. For a regulator or platform compliance officer aiming to:

\begin{itemize}[leftmargin=1.4em,itemsep=0.3em]
\item identify \emph{whom} to investigate, the lifecycle-and-conviction heuristic of \citet{gomezcram2026informed}---once category-conditioned and reproduced with released account labels---is the natural population-level filter;
\item quantify \emph{how much} information was extracted on \emph{which} specific contract, ILS$^{\mathrm{dl}}$ is the natural per-market measure;
\item establish \emph{whether} a specific trader actually possessed non-public information, neither methodology is sufficient---this is the proper domain of legal and regulatory investigation, which addresses questions that statistical inference cannot.
\end{itemize}

The layers stack. The Maduro / Venezuela cluster does not show that one approach is correct and the others wrong; it shows that each layer answers a question the others do not, and that for a high-profile case with public enforcement evidence, two of the three layers are observed and the third specifies the missing per-contract quantification that would complete the picture.

\section{A combined surveillance pipeline}
\label{sec:pipeline}

We sketch a three-stage surveillance pipeline that integrates the two academic methodologies under consideration. The pipeline is not a proposal for production deployment; we offer it as an organizing framework for thinking about how the two layers compose.

\subsection{Stage 1 --- category-conditioned account risk scoring}

Stage 1 runs on a periodic batch schedule (e.g., monthly) over the platform's complete on-chain trade history. The naive form of Stage 1---``run sign-randomization classification, output flagged accounts''---is, on the analysis of \Cref{sec:signrand}, insufficient as a surveillance layer. We sketch a stronger form that we will refer to as \emph{category-conditioned account risk scoring}.

The inputs to Stage 1 are:

\begin{itemize}[leftmargin=1.4em,itemsep=0.3em]
\item Per-account sign-randomization skill classification, computed separately within each target category (politics, sports, crypto, finance, etc.) rather than pooled at the platform level. This produces a per-account, per-category skill statistic that is mechanism-anchored: a sports-category skill is interpretable as injury-news / line-shading / arbitrage skill; a politics-category skill is interpretable in terms of public-news ingestion or, on the upper tail, suspected leak-driven trading. We note that category-conditioning trades interpretability for statistical power: an account that traded fifteen events overall but only two in geopolitics will have an underpowered within-category skill statistic for geopolitics, even if its pooled multi-category statistic was significant. Operationally, this means that the category-conditioned classifier will produce reliable per-category skill labels only for accounts that meet a within-category minimum-event threshold, and the population covered by Stage 1 will accordingly be smaller than the pooled-classification population.
\item Per-account lifecycle-and-conviction heuristic flags as in \Cref{sec:signrand-heuristic}, computed at the per-event level. This catches the one-shot informed-trader pattern that the skill classifier explicitly cannot reach.
\item Per-account composite anomaly score along the lines of \citet{mitts2026iran}, applied selectively to markets of regulatory interest (markets with sufficient volume and event-class membership).
\item Per-account context features: wallet age, funding-source provenance (CEX-deposit vs.\ on-chain mixer), known sybil-cluster membership, account participation in airdrop or promotional campaigns, position concentration within events of interest, and any known cross-platform identifiers.
\item Category-specific prior risk weights: a high skill score in a sports market is weighted differently from a high skill score in a geopolitical-event market, because the regulatory implications of skill in those domains differ. The weights are themselves a calibration parameter, and the methodology should support auditing and revision as ground-truth cases accumulate.
\end{itemize}

The output of Stage 1 is a per-account risk score that is mechanism-aware: an account that scores high on sports skill, has a wallet age of two years, and has funding entirely from a regulated CEX is treated differently from an account that scores high on geopolitics skill, was opened seven days before a high-stakes military event, and has funding from a privacy-preserving mixer. Accounts above a category-conditioned threshold are flagged forward to Stage 2; the cardinality of the flagged set is tunable by the threshold and by the category-specific weights.

Stage 1 alone produces high false-positive rates if used as a sole filter for surveillance action, because each input feature has its own false-positive distribution and category-conditioned aggregation does not eliminate this. It is a triage stage, not a terminal one. The naive ``run sign-randomization classifier and flag the $1{,}950$ heuristic insiders'' workflow is, in our reading, insufficient on three grounds: it pools categorically heterogeneous skill into a single number; it relies on a heuristic with unmeasured precision; and the underlying account labels would, for operational use, need to be reproduced and released, which has not yet been done by \citet{gomezcram2026informed}.

\subsection{Stage 2 --- per-market scoring on flagged-wallet markets}

Stage 2 operates on individual markets after a public event timestamp has been identified and before formal resolution. For each market in which one or more flagged accounts has taken a significant position---a threshold that is itself a calibration parameter, but in the simplest version, ``flagged account holds $\geq X\%$ of long-side notional''---Stage 2 computes ILS$^{\mathrm{dl}}$ against the current best estimate of $\Tevent$, together with short-window variants. Markets in which no flagged account holds a significant position are not scored, so the per-market computational cost is bounded by the number of flagged-account-active markets at any given time, which is small relative to the platform's total open-market count.

The output of Stage 2 is a per-market score gated by scope conditions. Markets failing the scope conditions (high opening-price extremes, trivial-resolution moves, anchor-insensitive scores) are not flagged forward. Markets passing the scope conditions and exhibiting $\ILS^{\mathrm{dl}}$ above a calibration threshold, in conjunction with a short-window variant above its threshold, are flagged forward. The companion empirical paper \citep{nechepurenko2026foresightflow_empirical} proposes a starting threshold of $\ILS^{\mathrm{dl}} > 0.25$ jointly with a short-window variant exceeding $0.10$, drawn from a single observation and explicitly preliminary.

\subsection{Stage 3 --- human review on jointly-flagged contracts}

Stage 3 applies human compliance review to the contracts surfaced by Stage 2: contracts where a flagged-account population holds significant position \emph{and} where the per-market ILS$^{\mathrm{dl}}$ exceeds threshold. The expected cardinality of Stage 3 is small enough to be tractable for human review on a continuing basis, by construction: the joint of Stage 1's wallet filter and Stage 2's market filter is much smaller than either alone.

Stage 3 is where the surveillance system interfaces with regulatory or law-enforcement processes; the academic methodologies do not directly produce indictments or platform suspensions, and we do not attempt to specify Stage 3 in this paper.

\subsection{Comparison to single-method approaches}

A single-method approach using only Stage 1 (wallet classification) is informative for population-economics characterization and identifies populations of interest, but it produces a flagged-account count---approximately $1{,}950$ for the heuristic insider category---that is too large for manual investigation as a routine surveillance practice and too imprecise for direct regulatory action. A single-method approach using only Stage 2 (per-market scoring) operates at the right unit for individual contract investigation but, without the wallet-level filter, applies to all open markets passing scope conditions and is computationally and operationally undirected.

The pipeline of the form Stage 1 $\to$ Stage 2 $\to$ Stage 3 inherits the methodological strengths of both layers and pushes the false-positive rate of each into a regime where joint flagging is plausibly tractable for human review. We emphasize again that this is an architectural sketch, not a calibrated proposal: each stage's threshold is an empirical question we do not address here, and the tractability claim is therefore qualitative.

\section{Open questions neither methodology addresses}
\label{sec:open}

We close by identifying three open questions that none of the three methodologies under consideration, in their current form, addresses.

\subsection{Resolution-criteria uncertainty}

The Iran-Apr30 case study \citep{nechepurenko2026foresightflow_empirical} illustrates a phenomenon that is methodologically distinct from informed trading: a market that broadly mispredicts its eventual resolution outcome because participants are uncertain about how the contract's resolution criteria will be applied to ambiguous events. In the Iran-Apr30 case, the market priced down to near-zero on consensus that the F-15E rescue operation did not constitute ``U.S.\ forces entering Iran,'' and was subsequently resolved YES by the UMA Optimistic Oracle.

Sign-randomization classifies the NO-buyers in this market as ``unskilled losers,'' which is misleading; they were in fact responding to publicly observable information about an ambiguous criterion, not exhibiting absence of skill. ILS$^{\mathrm{dl}}$ produces a coherent score against the article-derived $\Tevent$, but the score does not separate ``front-loading on informed prediction of the underlying event'' from ``front-loading on informed prediction of how the criterion would be interpreted''---if such a separation is even meaningful. A diagnostic that explicitly separates resolution-criteria uncertainty from informed front-loading would complement both wallet-level methods and per-market scoring; we are not aware of such a diagnostic in the present literature.

\subsection{Continuous pre-event trade collection}

Both methodologies are constrained, in their applicability to specific cases, by the retrospective availability of per-trade history. The companion empirical paper \citep{nechepurenko2026foresightflow_empirical} reports that Iran-cluster trade history was retrievable only from the resolution-settlement window onward, owing to The Graph subgraph indexer's coverage policy on low-volume markets at the time of indexing decision. \citet{gomezcram2026informed} have access to the full Polymarket transaction history because they query at population-scale retrospectively; per-market real-time access requires continuous collection from $\Topen$ for each open market in the target categories.

This is an infrastructure problem rather than a methodological one, but it is the binding constraint on both approaches' application to surveillance use cases. Continuous per-trade collection would unblock real-time wallet-level features (for sign-randomization-style methods) and enable wallet-level diagnostics on ILS$^{\mathrm{dl}}$-flagged markets. The methodological approaches under consideration depend on such infrastructure existing.

\subsection{Cross-market coordination}

Neither sign-randomization nor ILS$^{\mathrm{dl}}$ directly addresses coordination patterns across multiple related markets. Sign-randomization aggregates an account's history across all markets; it does not test whether an account's trades on Market $A$ are correlated with another account's trades on Market $B$ in ways suggestive of coordinated information flow. ILS$^{\mathrm{dl}}$ scores individual markets; it does not test whether high ILS$^{\mathrm{dl}}$ on Market $A$ is statistically correlated with high ILS$^{\mathrm{dl}}$ on Market $B$ for related underlying events.

The Iran-cluster wallet analysis in \citet{nechepurenko2026foresightflow_empirical}, although limited by trade-history availability, identified $332$ wallets active in both of two cluster markets in the resolution-settlement window. With pre-event trade collection in place, a cross-market coordination diagnostic would test whether such overlap is statistically anomalous prior to the underlying event, conditional on the events' related underlying causes. The Maduro/Venezuela cluster, with $7$ YES-resolved markets sharing a single underlying event (\emph{Operation Absolute Resolve}), is a natural test bed.

\section{Conclusion}
\label{sec:conclusion}

Multiple methodologically distinct approaches to detecting informed trading on Polymarket emerged in close succession in early 2026: a wallet-market composite anomaly screen \citep{mitts2026iran}, a wallet-level sign-randomization skill classifier together with a separate single-event lifecycle-and-conviction heuristic \citep{gomezcram2026informed}, and a market-level information-leakage score \citep{nechepurenko2026foresightflow_methodology, nechepurenko2026foresightflow_empirical}. The strongest organizing claim of this paper is that these are distinct \emph{layers} of detection rather than competing methods on a single layer. Sign-randomization, in particular, identifies persistent directional skill conditional on opportunity selection; this is a substantively different question from identifying insider trading in the regulatory sense. The Polymarket sample on which the methodology is evaluated combines economically heterogeneous categories (sports, politics, crypto, finance) with structurally different information technologies, so a platform-wide skill classification is mechanism-ambiguous and should be category-conditioned before deployment in a surveillance context. The lifecycle-and-conviction heuristic captures the canonical one-shot insider pattern that the skill classifier excludes by design, but its precision against an external ground-truth set has not been reported and its operational use in a surveillance pipeline requires both reproduction with released account labels and validation against known cases.

The Maduro / Venezuela cluster illustrates how the layers stack constructively. The lifecycle heuristic of \citet{gomezcram2026informed} flags three accounts whose realized profits collectively exceed \$$626{,}000$, with the lead account's PnL of \$$409{,}882$ matching the U.S.\ Department of Justice indictment's quoted Van Dyke profit of \$$409{,}881$ to the dollar---a striking face-valid alignment that establishes recall on at least this case while leaving precision on the full population an open question. Legal and regulatory investigation \citep{doj2026vandyke, cftc2026vandyke} addresses whether a specific named trader actually possessed non-public information, which is not a question that any statistical methodology can answer. ILS$^{\mathrm{dl}}$ would in principle quantify, contract by contract, how much information was front-loaded into each of the seven YES-resolved markets in the cluster before public observation of the underlying event; this per-contract decomposition is structurally absent from both \citet{gomezcram2026informed} components and from the DOJ indictment. On the present cluster, two of the three layers are empirically observed (lifecycle screening and legal investigation), and the third specifies the missing per-contract quantification that would complete the layered analysis.

Three open questions---resolution-criteria uncertainty, continuous pre-event trade collection, and cross-market coordination---remain outside the scope of the methodologies under consideration and define a research agenda for the next stage of the field. The combined surveillance pipeline sketched in \Cref{sec:pipeline} uses category-conditioned account risk scoring as Stage 1 (rather than a naive ``run sign-randomization classifier'' workflow), ILS$^{\mathrm{dl}}$ as Stage 2, and human compliance review as Stage 3. Each stage has known limitations; the value of the composition is precisely that the limitations of each stage are partially offset by the strengths of the others.

We close with a constructive note. The \citet{gomezcram2026informed} paper is, in our reading, a substantial contribution to the platform-economics literature on prediction markets, with a calibrated methodology, a striking skill-persistence empirical finding, and a documented case study aligned with subsequent enforcement action. The recommendations in this paper for category-conditioned decomposition, account-label release, and methodological separation between the skill classifier and the lifecycle heuristic are directed at the operationalization of the methodology for surveillance, not at the academic contribution of the paper. We expect that with such operationalization, sign-randomization-style classification and ILS$^{\mathrm{dl}}$-style per-market scoring will compose into a substantially stronger surveillance capacity than either methodology alone provides.

\section*{Revision note (v2)}

This v2 revision is a coordinated update with the companion methodology and empirical papers \citep{nechepurenko2026foresightflow_methodology, nechepurenko2026foresightflow_empirical}. Two findings from the platform data audit motivate the revision: (i) a category-labeling correction reclassifying $15{,}542$ esports markets (predominantly Counter-Strike) out of the military / geopolitical category, and (ii) an expansion of the Tier-3 event-timestamp recovery sample used in the companion empirical paper from the v1 budget-capped subset ($n = 9$) to the full available Tier-3 population ($n = 18$) for military / geopolitical markets. The v1 hazard estimate ($\hat\lambda = 0.306$, half-life $2.3$\,d) lies inside the v2 95\% confidence interval $[0.143, 0.365]$, so the two estimates are statistically consistent; the v2 point estimate is $\hat\lambda = 0.241$ with half-life $2.9$\,d and KS adequacy $p = 0.426$.

The methodological comparison developed in this paper is unaffected by either finding. The taxonomy of detection layers (\Cref{tab:taxonomy}), the analysis of category heterogeneity (\Cref{sec:signrand}), the separation between the sign-randomization skill classifier and the lifecycle-and-conviction heuristic, and the layered reading of the U.S.--Venezuela operation cluster (\Cref{sec:convergence}) all stand as written in v1. Cross-references to the companion empirical paper's hazard estimates now resolve to the v2 values via the updated bibliography. Three accompanying datasets are released openly with this revision: \emph{polymarket-tnews-tevent-recovery-v1}, \emph{polymarket-hazard-rates-v1}, and \emph{polymarket-ils-corpus-v1}, all at \url{https://github.com/ForesightFlow/datasets}.

\section*{Generative AI Disclosure}

In preparing this manuscript, the author used Anthropic's Claude Opus 4.7 for copy-editing, structural review against an external referee report, and rendering of tables from numerical data. All methodology, analysis, and conclusions are the author's own; the author reviewed and edited all AI-generated content and takes full responsibility for the final manuscript.

\bibliographystyle{plainnat}
\bibliography{refs}

\end{document}